\documentclass[final,5p,times,twocolumn]{elsarticle}

\usepackage{amsmath}
\usepackage{amssymb}

\biboptions{numbers,sort&compress}

\newcommand{\bg}{\bar{\gamma}}

\newcommand{\bh}{\bar{h}}
\newcommand{\barf}{\bar{f}}

\newcommand{\bpsi}{\bar{\psi}}
\newcommand{\bA}{\bar{A}_0}

\journal{Physics Letters B}

\begin{document}

\begin{frontmatter}

\title{Effect of scalar field mass on gravitating charged scalar solitons and black holes in a cavity}

\author{Supakchai Ponglertsakul}
\ead{supakchai.p@gmail.com}

\author{Elizabeth Winstanley}
\ead{E.Winstanley@sheffield.ac.uk}

\address{Consortium for Fundamental Physics, School of Mathematics and Statistics, University of Sheffield,\\
Hicks Building, Hounsfield Road, Sheffield S3 7RH, United Kingdom}

\date{\today}

\begin{abstract}
We study soliton and black hole solutions of Einstein charged scalar field theory in cavity. We examine the effect of introducing a scalar field mass on static, spherically symmetric solutions of the field equations. We focus particularly on the spaces of soliton and black hole solutions, as well as studying their stability under linear, spherically symmetric perturbations of the metric, electromagnetic field, and scalar field.
\end{abstract}


\end{frontmatter}

\section{Introduction}
\label{sec:intro}

In the phenomenon of charge superradiance, a classical charged scalar field wave incident on a Reissner-Nordstr\"om black hole is scattered with a reflection coefficient of greater than unity if the frequency, $\omega $, of the wave satisfies the inequality \cite{Bekenstein:1973mi}
\begin{equation}
0< \omega < q\Phi _{h},
\label{eq:chargeSR}
\end{equation}
where $q$ is the charge of the scalar field and $\Phi _{h}$ is the electrostatic potential at the event horizon of the black hole.
By this process, the charged scalar field wave extracts some of the electrostatic energy of the black hole.
If a charged scalar field wave satisfying (\ref{eq:chargeSR}) is trapped near the event horizon by a reflecting mirror of radius $r_{m}$, the wave can scatter repeatedly off the black hole, and is amplified each time it is reflected.
This can lead to an instability (the ``charged black hole bomb'') where the amplitude of the wave grows exponentially with time \cite{Herdeiro:2013pia, Degollado:2013bha, Hod:2013fvl, Hod:2016nsr}, providing the scalar field charge $q$ and mass $\mu $ satisfy the inequality \cite{Hod:2016nsr}
\begin{equation}
\frac {q}{\mu } > {\sqrt {\frac {\frac {r_{m}}{r_{-}}-1}{\frac {r_{m}}{r_{+}}-1}}}>1,
\label{eq:instabcond}
\end{equation}
where $r_{+}$ and $r_{-}$ are, respectively, the radius of the event horizon and inner horizon of the black hole.
The inequality (\ref{eq:instabcond}) ensures that the area of the event horizon increases as the scalar field evolves \cite{Herdeiro:2013pia}, and implies that for fixed $q$ and $\mu $, the mirror radius $r_{m}$ must be sufficiently large for an instability to occur.
Physically, the scalar field wave must extract more charge than mass from the black hole, so that the black hole evolves away from extremality.

What is the ultimate fate of this charged black hole bomb instability?
To answer this question, it is necessary to go beyond the test-field limit and consider the back-reaction of the charged scalar field on the black hole geometry.
Recently, we studied static, spherically symmetric, black hole \cite{Dolan:2015dha} and soliton \cite{Ponglertsakul:2016wae} solutions of Einstein charged scalar field theory in a cavity, in the case where the scalar field mass $\mu $ is set equal to zero.
For both soliton and black hole solutions, the scalar field vanishes on the mirror.
We examined the stability of these charged-scalar solitons and black holes by considering linear, spherically symmetric, perturbations of the metric, electromagnetic field, and massless charged scalar field.
In the black hole case \cite{Dolan:2015dha}, we found that if the scalar field has no zeros between the event horizon and mirror, then the black holes appear to be stable.
On the other hand, if the scalar field vanishes inside the mirror then the system is unstable.
The situation for solitons is more complex \cite{Ponglertsakul:2016wae}.
Even if the scalar field has no zeros inside the mirror, there are some solitons which are unstable.
The unstable solitons have small mirror radius and large values of the electrostatic potential at the origin.

In \cite{Dolan:2015dha} we conjectured that the stable black holes with charged scalar field hair could be possible end-points of the charged black hole bomb
instability.
This conjecture has been tested recently \cite{Sanchis-Gual:2015lje, Sanchis-Gual:2016tcm} by evolving the fully coupled, time-dependent, spherically symmetric, Einstein-Maxwell-Klein-Gordon equations in a cavity.
Starting from a Reissner-Nordstr\"om black hole in a cavity with a small charged scalar field perturbation, the system evolved to a hairy black hole in which some of the charge of the original black hole was transferred to the scalar field.

For a massless charged scalar field, the work of \cite{Sanchis-Gual:2016tcm} confirms our conjecture in \cite{Dolan:2015dha} - the ultimate fate of the charged black hole bomb is an equilibrium black hole with scalar field hair.  However, in \cite{Sanchis-Gual:2015lje, Sanchis-Gual:2016tcm} a massive charged scalar field is also considered.
In this paper we therefore study the effect of introducing a scalar field mass on the soliton and black hole solutions found in \cite{Dolan:2015dha, Ponglertsakul:2016wae}.
Our aim is to examine whether the end-points of the charged black hole bomb instability found in \cite{Sanchis-Gual:2015lje, Sanchis-Gual:2016tcm} correspond to stable equilibrium solutions of the Einstein-Maxwell-Klein-Gordon equations.

To this end, we begin in section \ref{sec:eqns} by introducing Einstein massive charged scalar field theory.
We study numerical soliton and black hole solutions of the static, spherically symmetric field equations in section \ref{sec:sols}, paying particular attention to the effect of the scalar field mass on the phase space of solutions.
The stability of the solutions is investigated in section \ref{sec:stab}, before our conclusions are presented in section \ref{sec:conc}.

\section{Einstein massive charged scalar field theory}
\label{sec:eqns}

We consider a self-gravitating massive charged scalar field coupled to gravity and an electromagnetic field, and described by the action
\begin{equation}
S = \frac {1}{2} \int {\sqrt {-g}} \, d^{4}x \left[ R - \frac {1}{2} F_{ab}F^{ab} - g^{ab} D^{*}_{(a }\Phi^{*} D_{b )}\Phi - \mu ^{2} \Phi ^{*}\Phi \right]
\label{eq:action}
\end{equation}
where $g$ is the metric determinant, $R$ the Ricci scalar, $F_{ab}=\nabla _{a}A_{b}-\nabla _{b}A_{a}$ is the electromagnetic field (with electromagnetic potential $A_{a}$), $\Phi $ is the complex scalar field, $\Phi ^{*}$ its complex conjugate and $D_{a} = \nabla _{a}-iqA_{a}$ with $\nabla _{a}$ the usual space-time covariant derivative.  Round brackets in subscripts denote symmetrization of tensor indices. The scalar field charge is $q$ and $\mu $ is the scalar field mass.  We use units in which $8\pi G=1=c$ and metric signature $(-,+,+,+)$.

Varying the action (\ref{eq:action}) gives the Einstein-Maxwell-Klein-Gordon equations
\begin{equation}
G_{ab}  =  T_{ab}^{F}+T_{ab}^{\Phi },
\qquad
\nabla _{a}F^{ab} = J^{b},
\qquad
D_{a}D^{a} \Phi - \mu ^{2} \Phi =0,
\label{eq:fieldeqns}
\end{equation}
where the stress-energy tensor $T_{ab}=T_{ab}^{F}+T_{ab}^{\Phi }$ is given by
\begin{align}
T_{ab}^{F}  = & F_{ac}F_{b}{}^{c} - \frac {1}{4} g_{ab} F_{cd}F^{cd} ,
\nonumber \\
T_{ab}^{\Phi }  = & D^{*}_{(a}\Phi ^{*}D_{b)}\Phi - \frac {1}{2}g_{ab} \left[ g^{cd} D^{*}_{(c}\Phi ^{*}D_{d)}\Phi + \mu ^{2}\Phi ^{*}\Phi \right] ,
\label{eq:Tmunu}
\end{align}
and the current $J^{a}$ is
\begin{equation}
J^{a} = \frac {iq}{2} \left[ \Phi ^{*} D^{a}\Phi - \Phi \left( D^{a}\Phi \right) ^{*} \right] .
\end{equation}

We consider static, spherically symmetric, solitons and black holes with metric ansatz
\begin{equation}
ds^{2}=-f(r)h(r)dt^{2} + f^{-1}(r) dr^{2} + r^{2}\left[ d\theta ^{2} + \sin ^{2} \theta \, d\varphi ^{2} \right] ,
\label{eq:metric}
\end{equation}
where the metric functions $f$ and $h$ depend only on the radial coordinate $r$. It is useful to define an additional metric function $m(r)$ by
\begin{equation}
f(r) = 1-\frac {2m(r)}{r}.
\label{eq:mdef}
\end{equation}
By a suitable choice of gauge (see \cite{Dolan:2015dha, Ponglertsakul:2016wae} for details), we can take the scalar field $\Phi =\phi (r)$ to be real and depend only on $r$.
The electromagnetic gauge potential has a single non-zero component which depends only on $r$, namely $A_{\mu } = \left[ A_{0}(r),0,0,0 \right]$.
Defining a new quantity $E=A_{0}'$, the static field equations (\ref{eq:fieldeqns}) generalize those in \cite{Dolan:2015dha, Ponglertsakul:2016wae} to include a nonzero scalar field mass and take the form
\begin{subequations}
\label{eq:static}
\begin{align}
h' &= r\left(qA_0\phi f^{-1}\right)^2 + rh\phi'^2,  \\
E^2 + \mu^2 h \phi^2 &= -\frac {2}{r} \left[ f'h + \frac{1}{2}fh' + \frac{h}{r}\left(f-1\right) \right] ,
\\
0 & = fA_0'' + \left(\frac{2f}{r} - \frac{fh'}{2h}  \right)A_0' - q^2\phi^2A_0 , \\
0 & =
f\phi'' + \left(\frac{2f}{r} + f' + \frac{fh'}{2h}    \right)\phi' + \left(\frac{q^2A_0^2}{fh} - \mu^2\right)\phi .
\end{align}
\end{subequations}

\section{Soliton and black hole solutions}
\label{sec:sols}

We now consider soliton and black hole solutions of the static field equations (\ref{eq:static}). In both cases we have a mirror at radius $r_{m}$, on which the scalar field must vanish, so that $\phi (r_{m})=0$.
As in \cite{Ponglertsakul:2016wae}, here we consider only solutions where the scalar field has its first zero on the mirror, since it is shown in \cite{Dolan:2015dha} that black hole solutions for which the scalar field has its second zero on the mirror are linearly unstable.

\subsection{Solitons}
\label{sec:soliton}

In order for all physical quantities to be regular at the origin, the field variables have the following expansions for small $r$:
\begin{align}
m &= \left(\frac{\phi_0^2\left[ a_0^2q^2 + h_0\mu^2\right] }{12h_0}\right)r^3 + O(r^5), \nonumber \\
h &= h_0 + \left(\frac{q^2a_0^2\phi_0^2}{2}\right)r^2 + O(r^4), \nonumber \\
A_0 &= a_0 + \left(\frac{a_0q^2\phi_0^2}{6}\right)r^2 + O(r^4), \nonumber \\
\phi &= \phi_0 - \left(\frac{\phi_0\left[ a_0^2 q^2 - h_0\mu^2 \right] }{6h_0}\right)r^2 + O(r^4),
\label{eq:origin}
\end{align}
where $\phi _{0}$, $a_{0}$ and $h_{0}$ are arbitrary constants. By rescaling the time coordinate (see \cite{Ponglertsakul:2016wae} for details), we can set $h_{0}=1$ without loss of generality.
A length rescaling \cite{Ponglertsakul:2016wae} can then be used to fix the scalar field charge $q=0.1$.
For each value of the scalar field mass $\mu $, soliton solutions are then parameterized by the two quantities $a_{0}$ and $\phi _{0}$.

\begin{figure}
\begin{center}
\includegraphics[width=0.9\columnwidth]{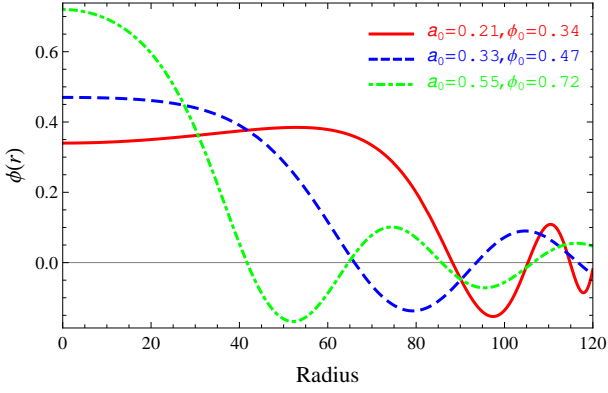}
\end{center}
\caption{Scalar field profiles for some typical soliton solutions with scalar field charge $q=0.1$ and mass $\mu =0.03$.}
\label{fig:solex}
\end{figure}

\begin{figure}
\begin{center}
\includegraphics[width=0.75\columnwidth]{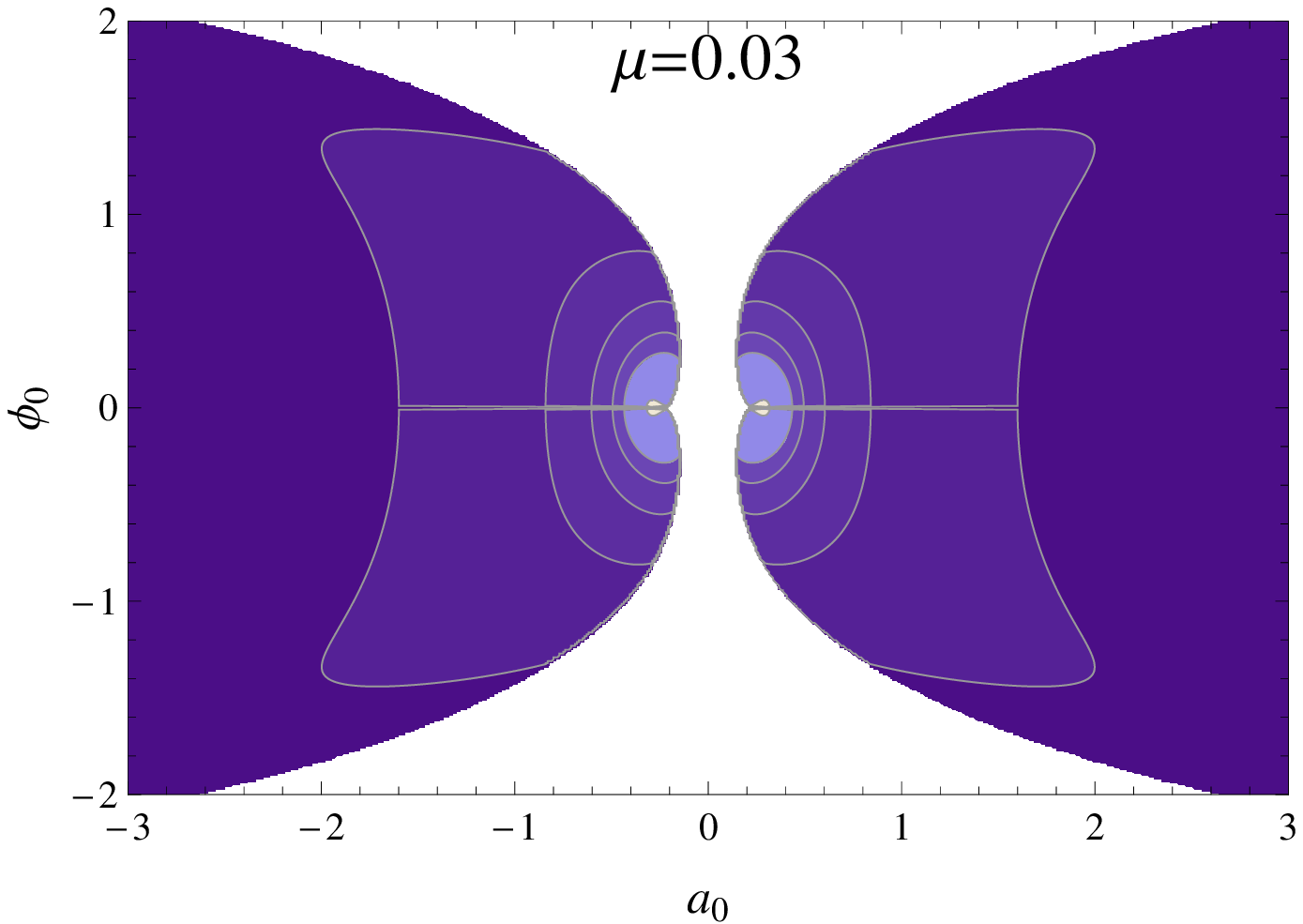}\\
\includegraphics[width=0.75\columnwidth]{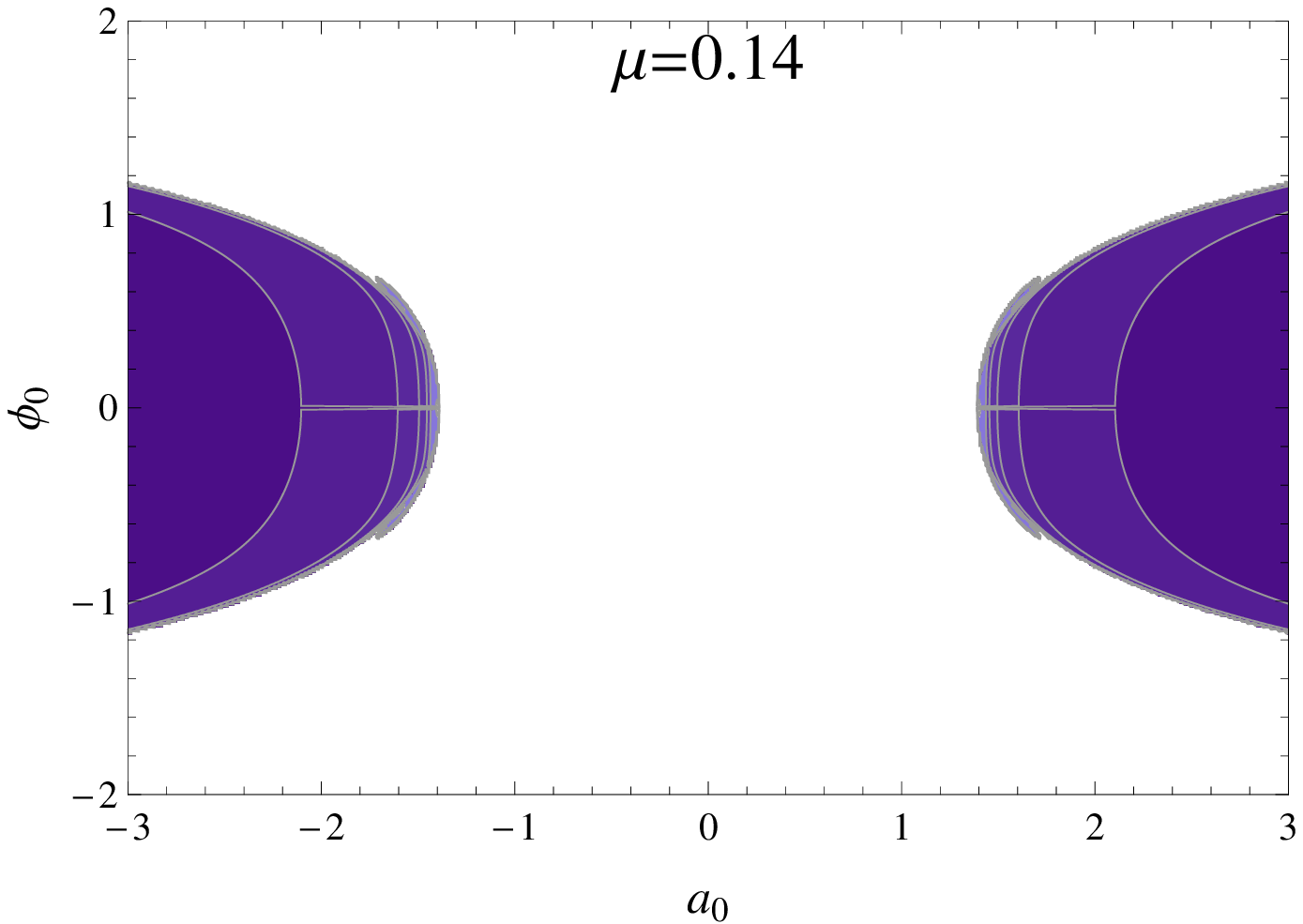}\\
\includegraphics[width=0.75\columnwidth]{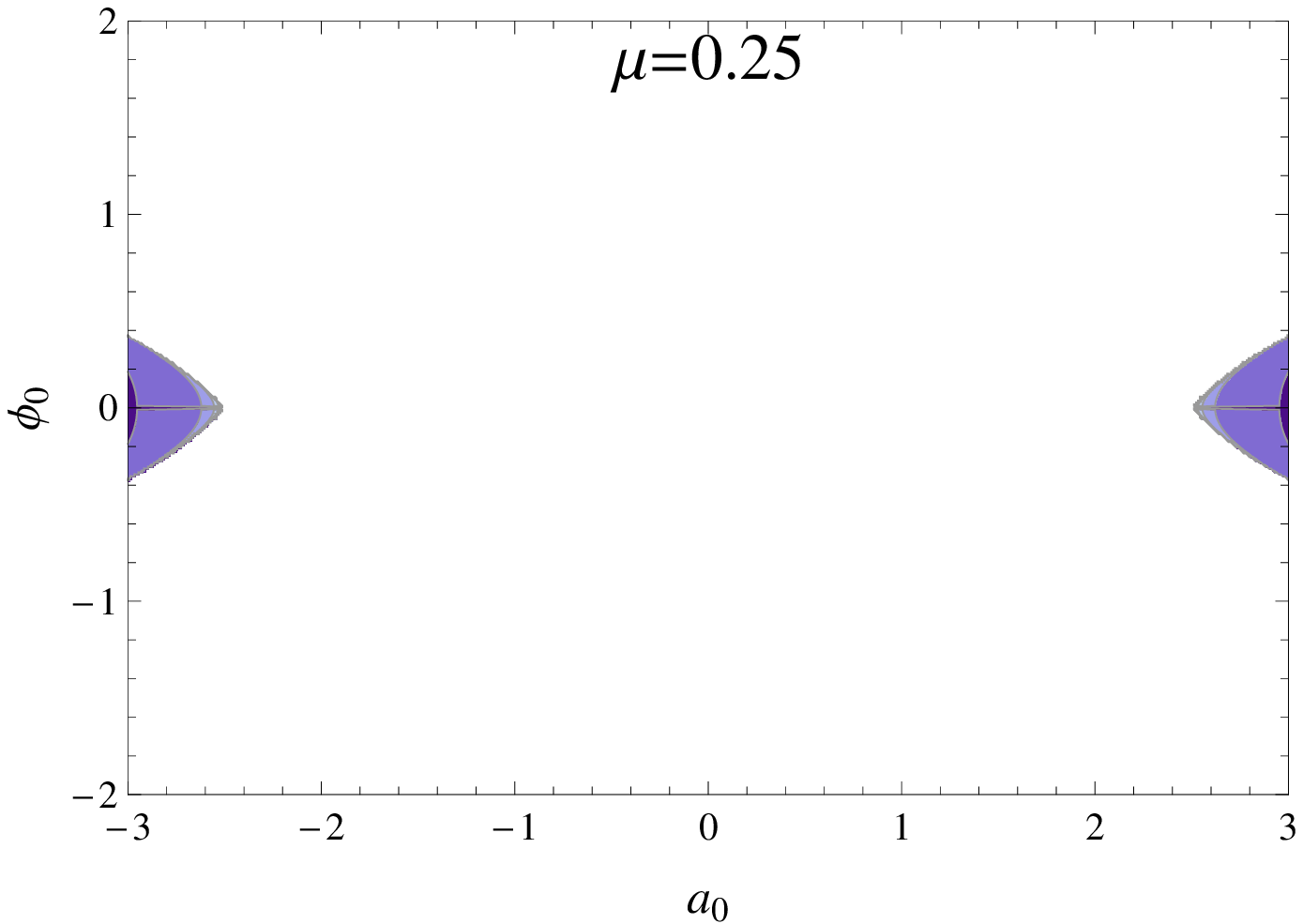}
\end{center}
\caption{Portions of the phase spaces of soliton solutions with scalar field charge $q=0.1$ and three values of the scalar field mass $\mu $.
Shaded regions indicate where solutions exist.  The curves are contours at constant mirror radius $r_{m}=20$, $40$, $60$, $80$, $100$ and $300$.
The darkest regions have $r_{m}<20$; for the lightest regions, the mirror radius $r_{m}>300$.}
\label{fig:solphase}
\end{figure}

Scalar field profiles for some typical soliton solutions are shown in figure \ref{fig:solex}.
From the expansions (\ref{eq:origin}), it can be seen that if the scalar field mass vanishes, $\mu =0$, and $\phi _{0}>0$ then close to the origin the scalar field is decreasing \cite{Ponglertsakul:2016wae}. This is no longer necessarily the case when $\mu >0$.  For $\phi _{0}>0$ and $h_{0}=1$, if
$\left| a_{0} \right| >\mu /q$ then the scalar field is decreasing close to the origin, and, for the numerical solutions investigated, it monotonically decreases to zero on the mirror.
If $\left| a_{0} \right|  <\mu /q$ then the scalar field is increasing close to the origin and must therefore have a maximum before decreasing to zero on the mirror.
This behaviour can be seen in figure \ref{fig:solex}.

We find that the phase space of solitons depends on the scalar field mass $\mu $, see figure \ref{fig:solphase}.  As in the massless case \cite{Ponglertsakul:2016wae}, for nonzero $\mu $ there appears to be no upper bound on the value of $\left| a_{0} \right| $ for which there are soliton solutions; accordingly only a portion of the phase space is shown in figure \ref{fig:solphase}.
When $\mu =0$, in \cite{Ponglertsakul:2016wae} we found solitons for $\left| a_{0} \right| $ arbitrarily small (but nonzero).  However, when $\mu >0$, we find that solitons exist only for $\left| a_{0} \right| $ above some lower bound, which increases as $\mu $ increases.
If $\phi _{0}>0$ and $\left| a_{0} \right| $ is too small, then the scalar field is increasing sufficiently rapidly close to the origin that it is unable to decrease to zero before either the metric function $f(r)$ has a zero or the solution becomes singular.

The other interesting feature in figure \ref{fig:solphase} is the existence of solitons with $\mu > q$. For such values of the scalar field mass, there is no charged black hole bomb instability in the test-field limit (\ref{eq:instabcond}).  We therefore now explore whether there are also black hole solutions when $\mu >q$.

\begin{figure}
\begin{center}
\includegraphics[width=0.9\columnwidth]{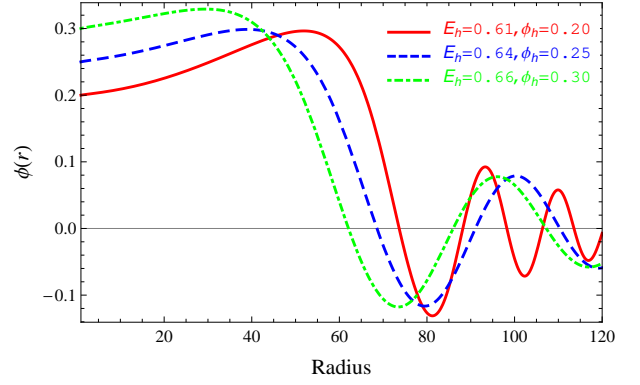}
\end{center}
\caption{Scalar field profiles for some typical black hole solutions with event horizon radius $r_{h}=1$, scalar field charge $q=0.1$ and mass $\mu = 0.07$.}
\label{fig:BHex}
\end{figure}

\begin{figure*}
\begin{center}
\begin{tabular}{ccc}
\includegraphics[width=0.6\columnwidth]{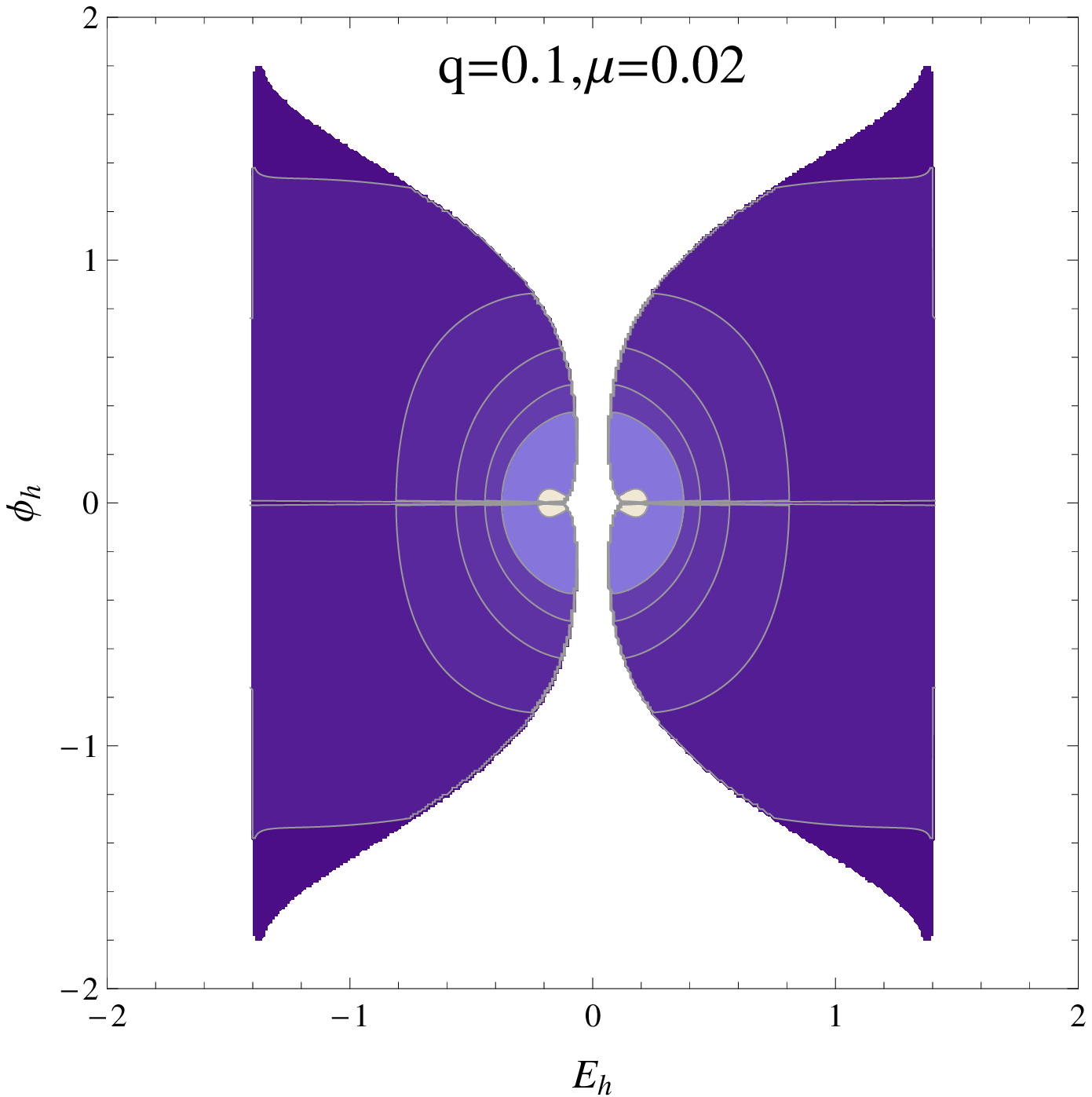}
\includegraphics[width=0.6\columnwidth]{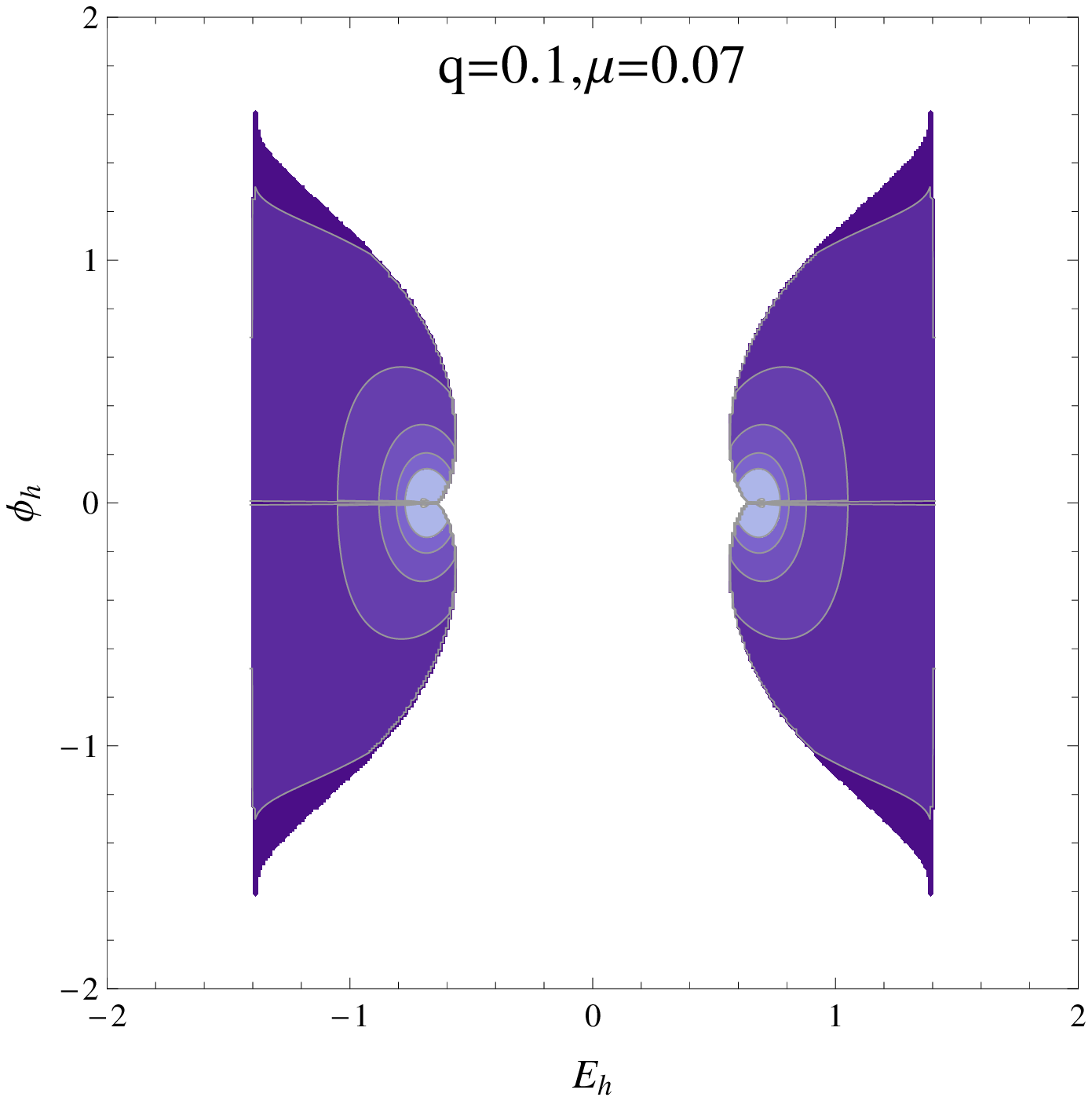}
\includegraphics[width=0.6\columnwidth]{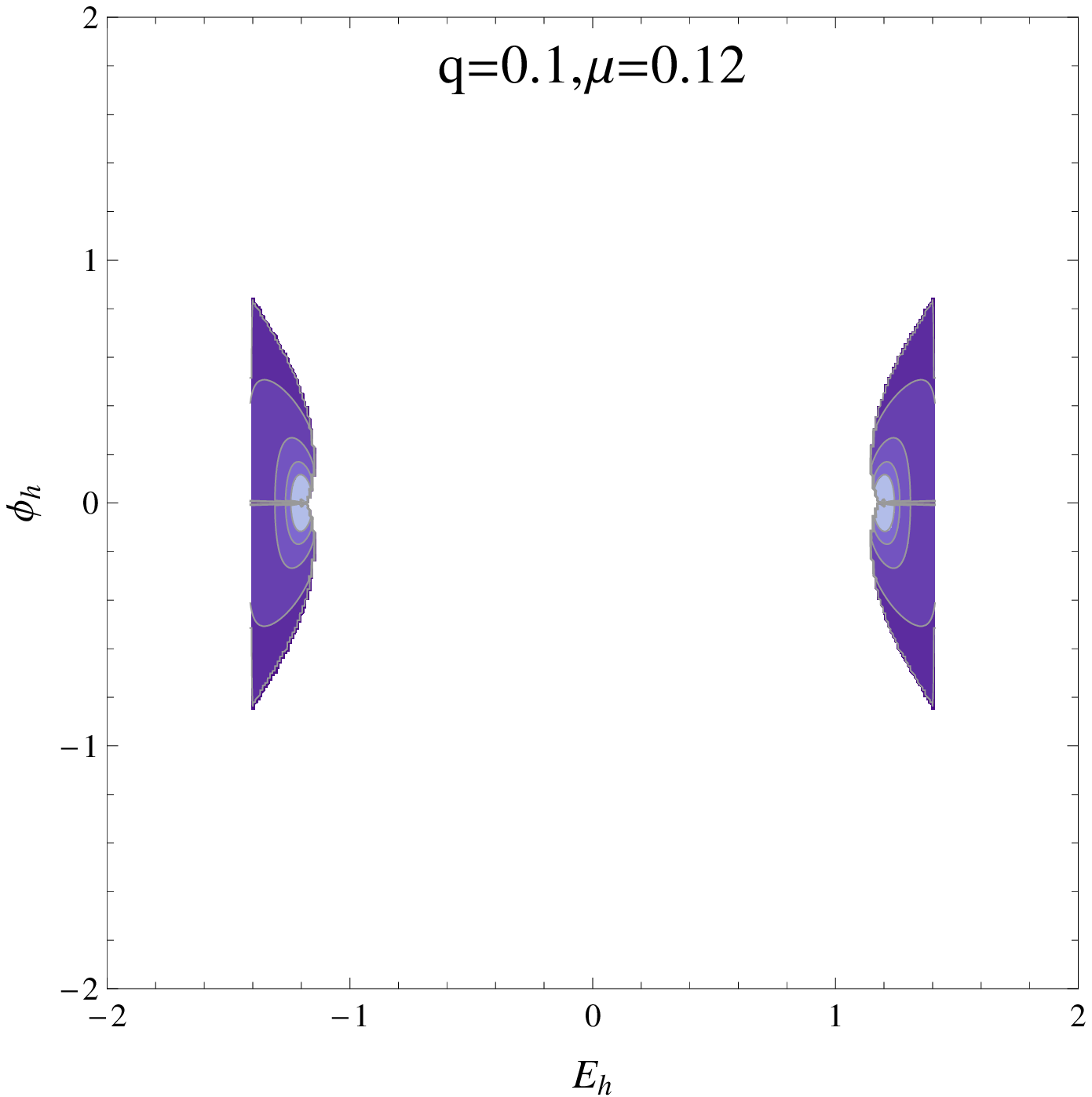}\\
\includegraphics[width=0.6\columnwidth]{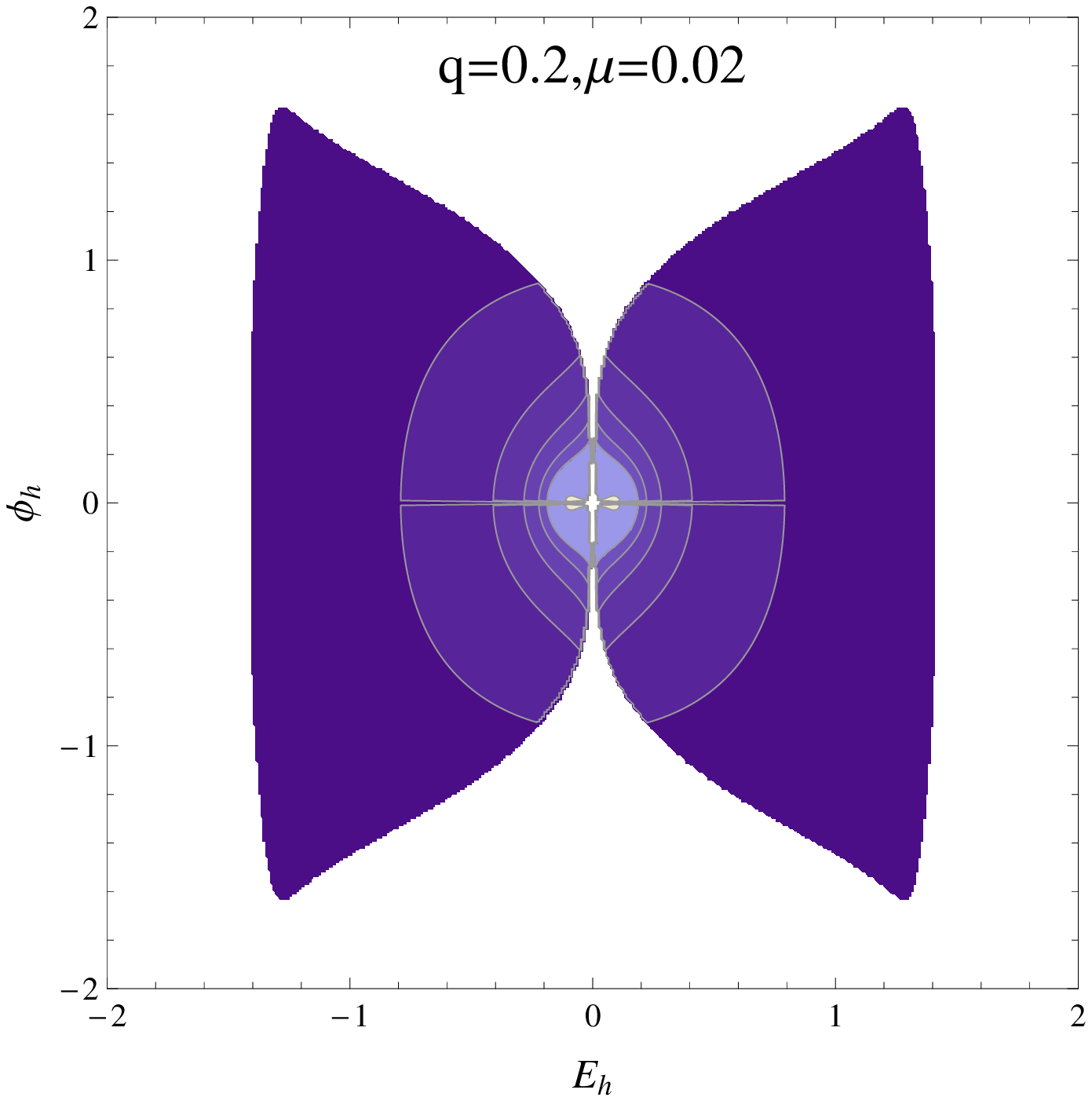}
\includegraphics[width=0.6\columnwidth]{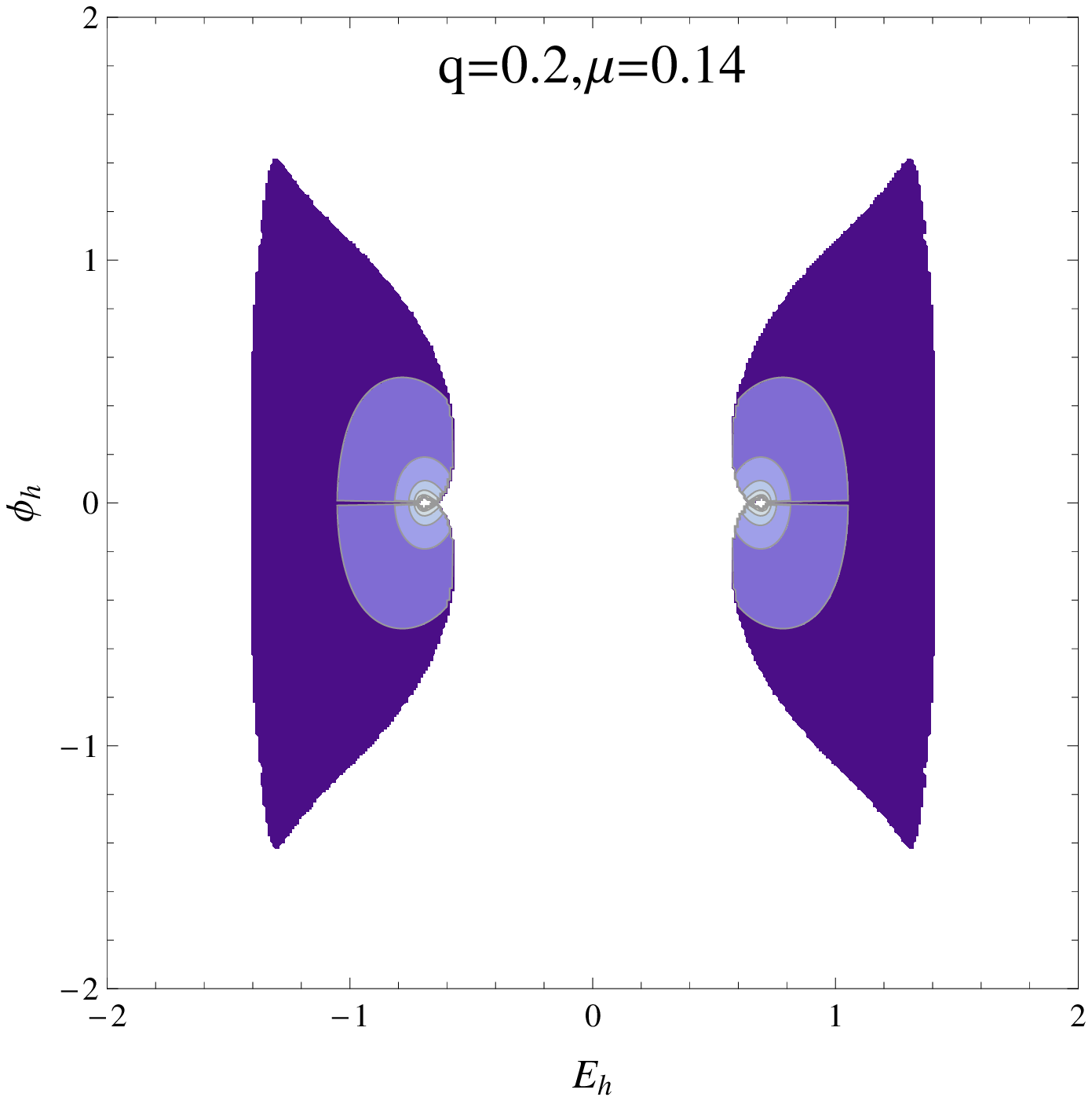}
\includegraphics[width=0.6\columnwidth]{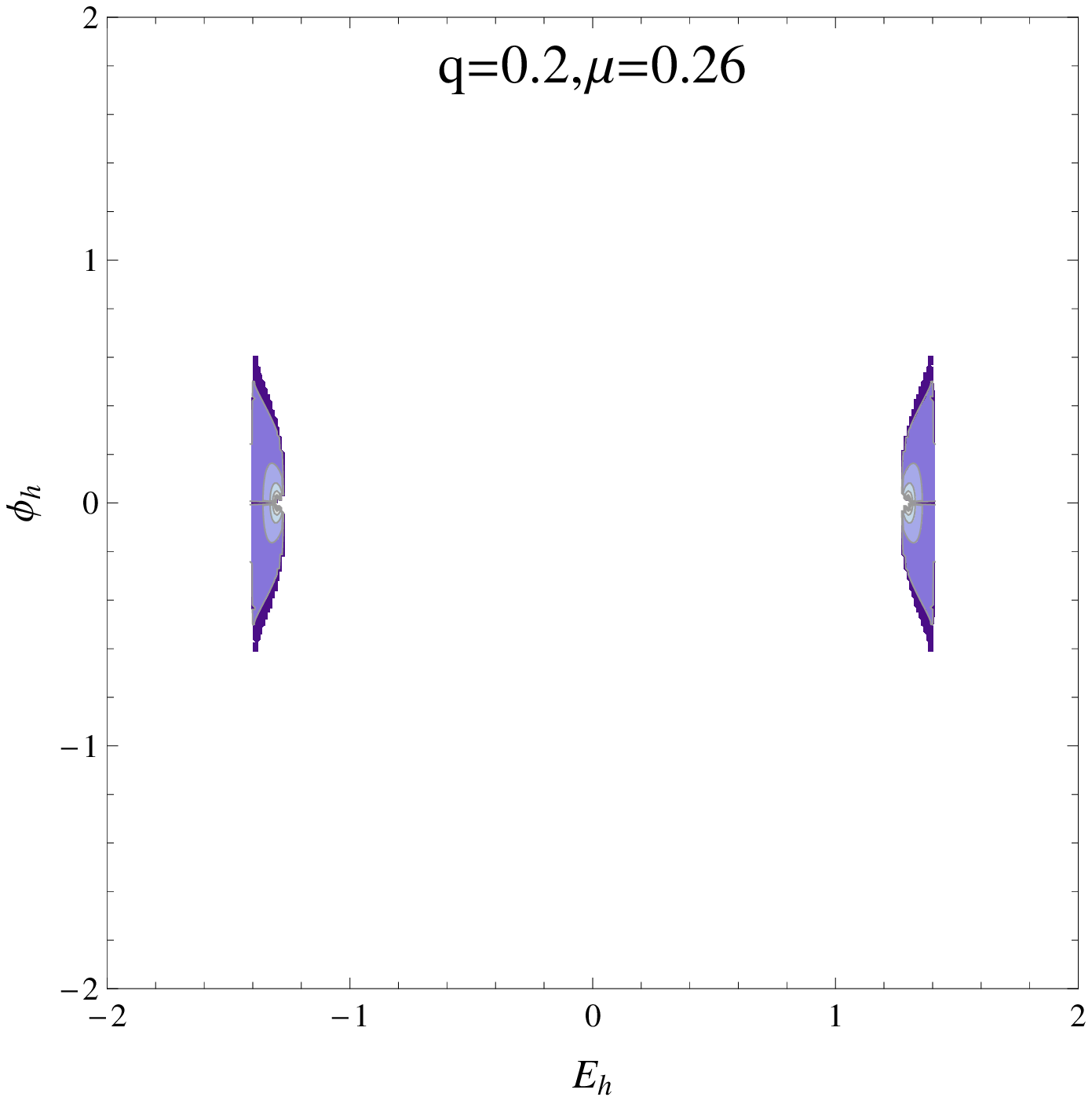}\\
\includegraphics[width=0.6\columnwidth]{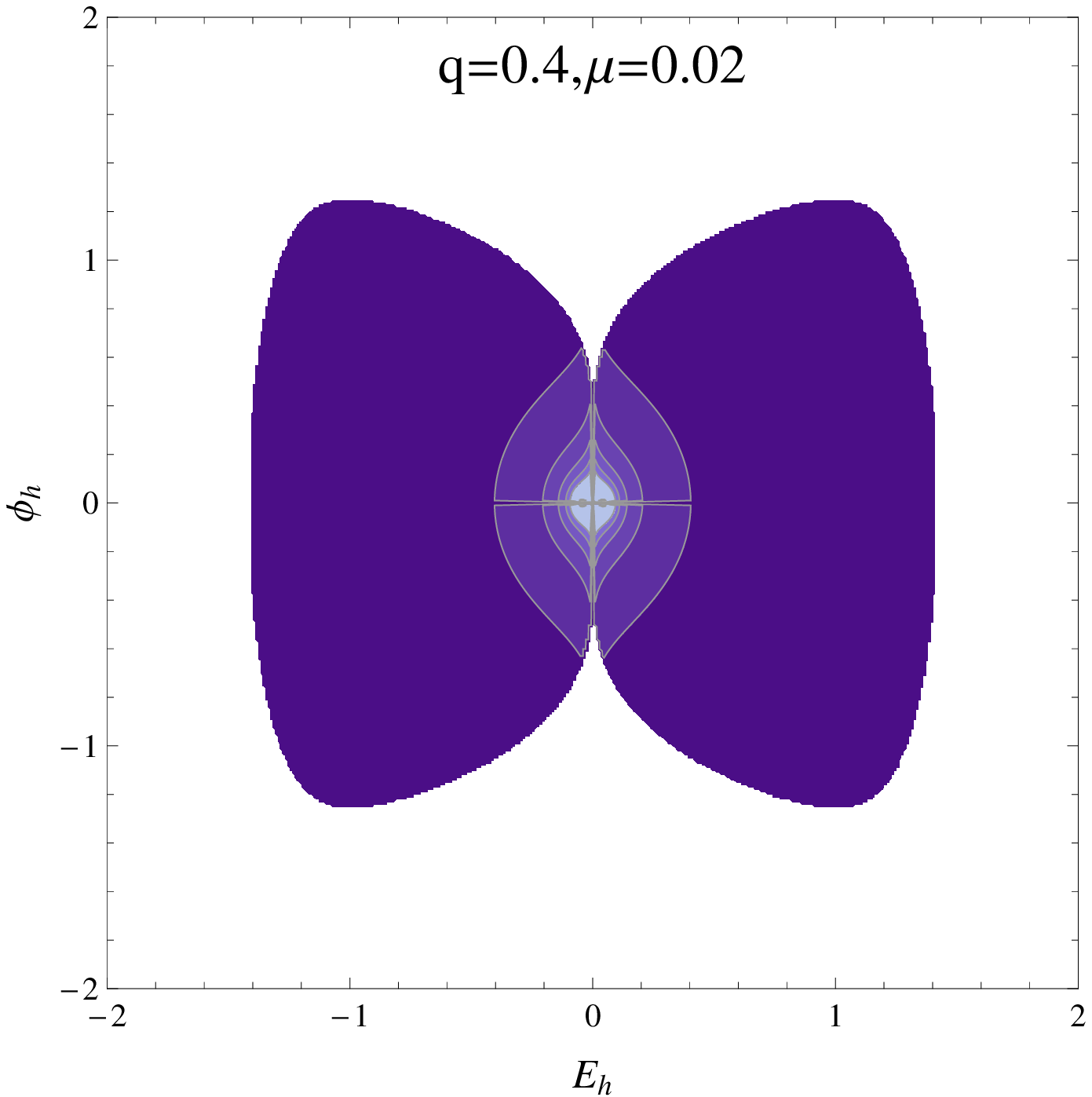}
\includegraphics[width=0.6\columnwidth]{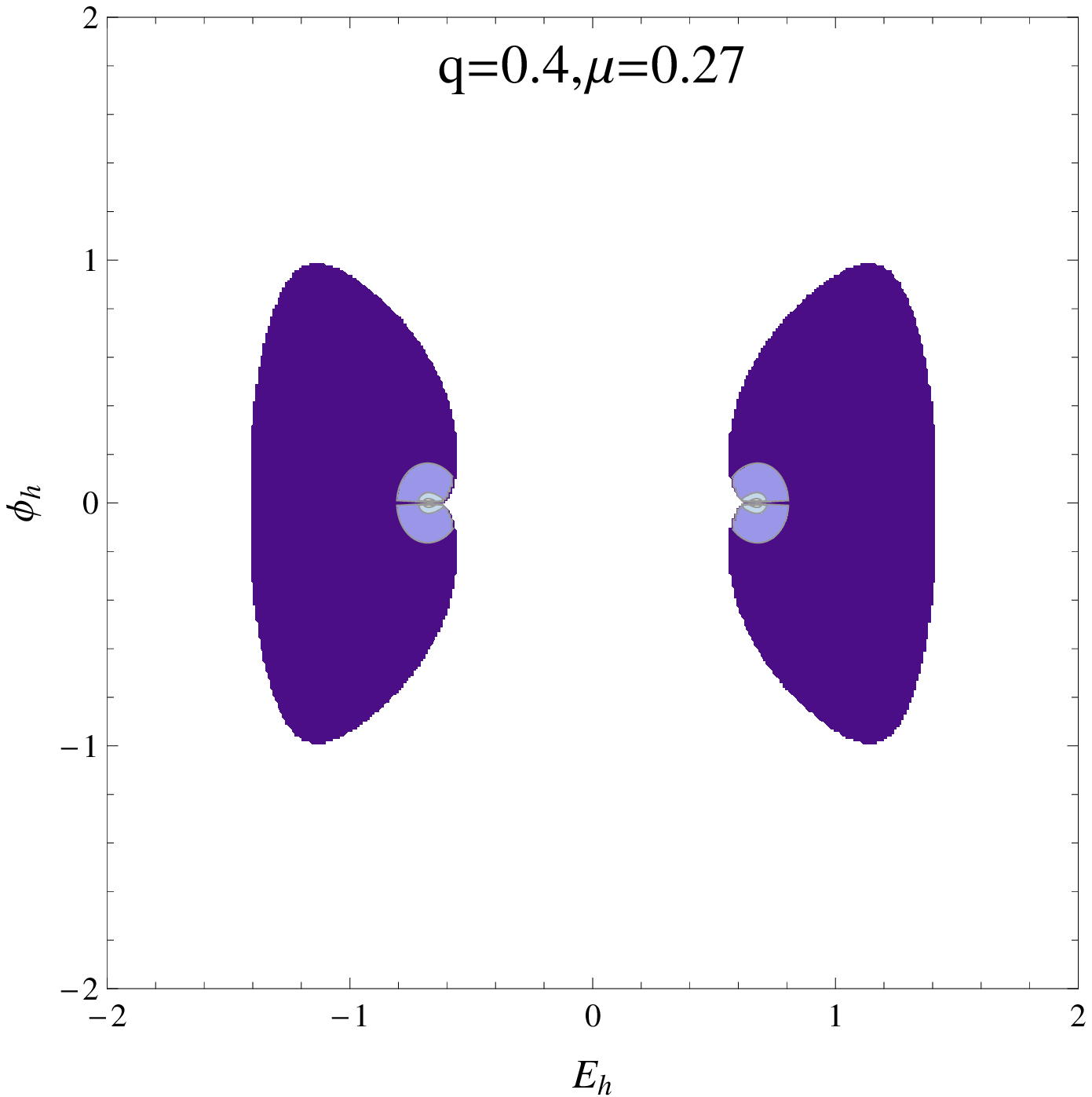}
\includegraphics[width=0.6\columnwidth]{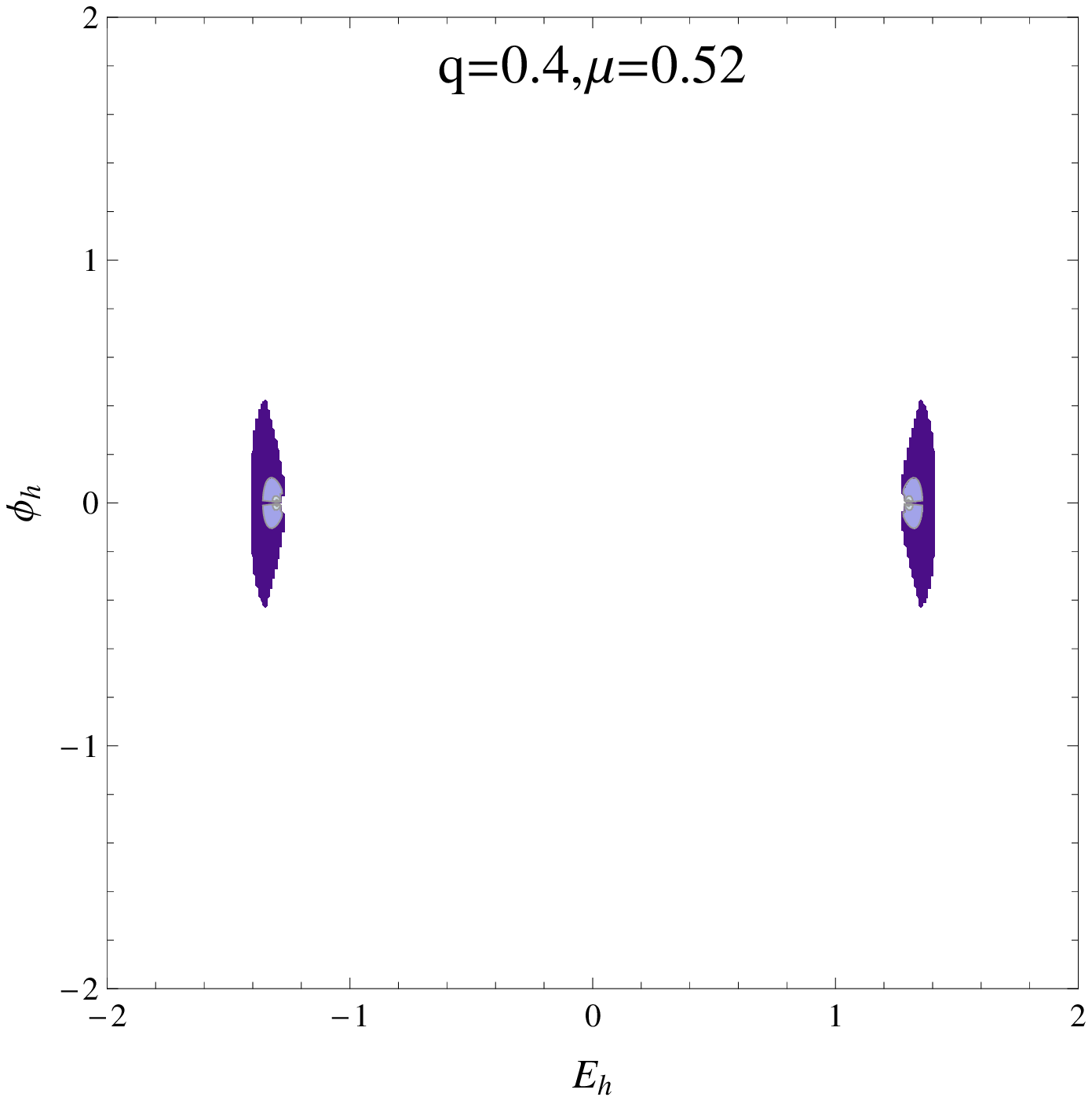}\\
\includegraphics[width=0.6\columnwidth]{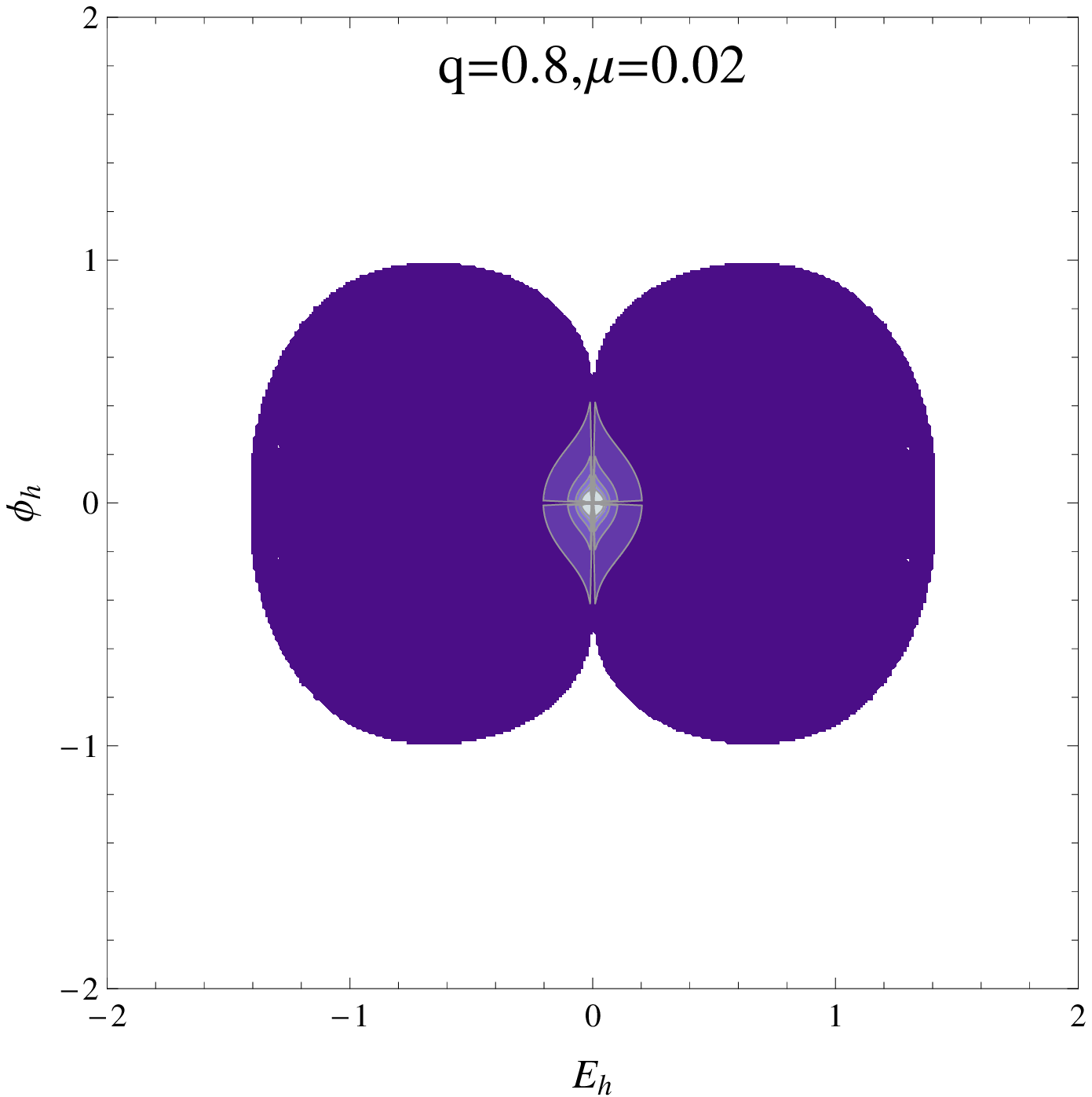}
\includegraphics[width=0.6\columnwidth]{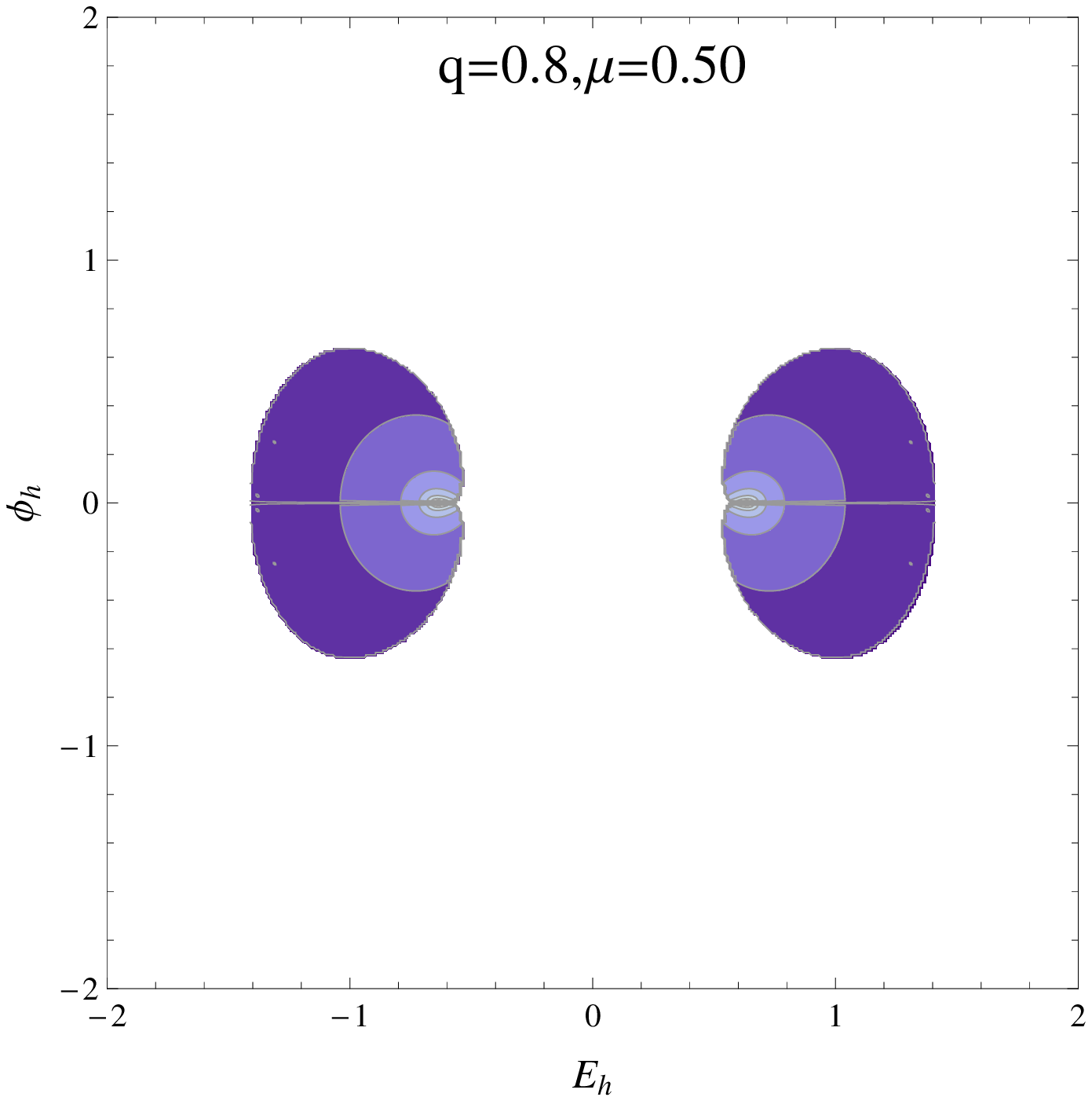}
\includegraphics[width=0.6\columnwidth]{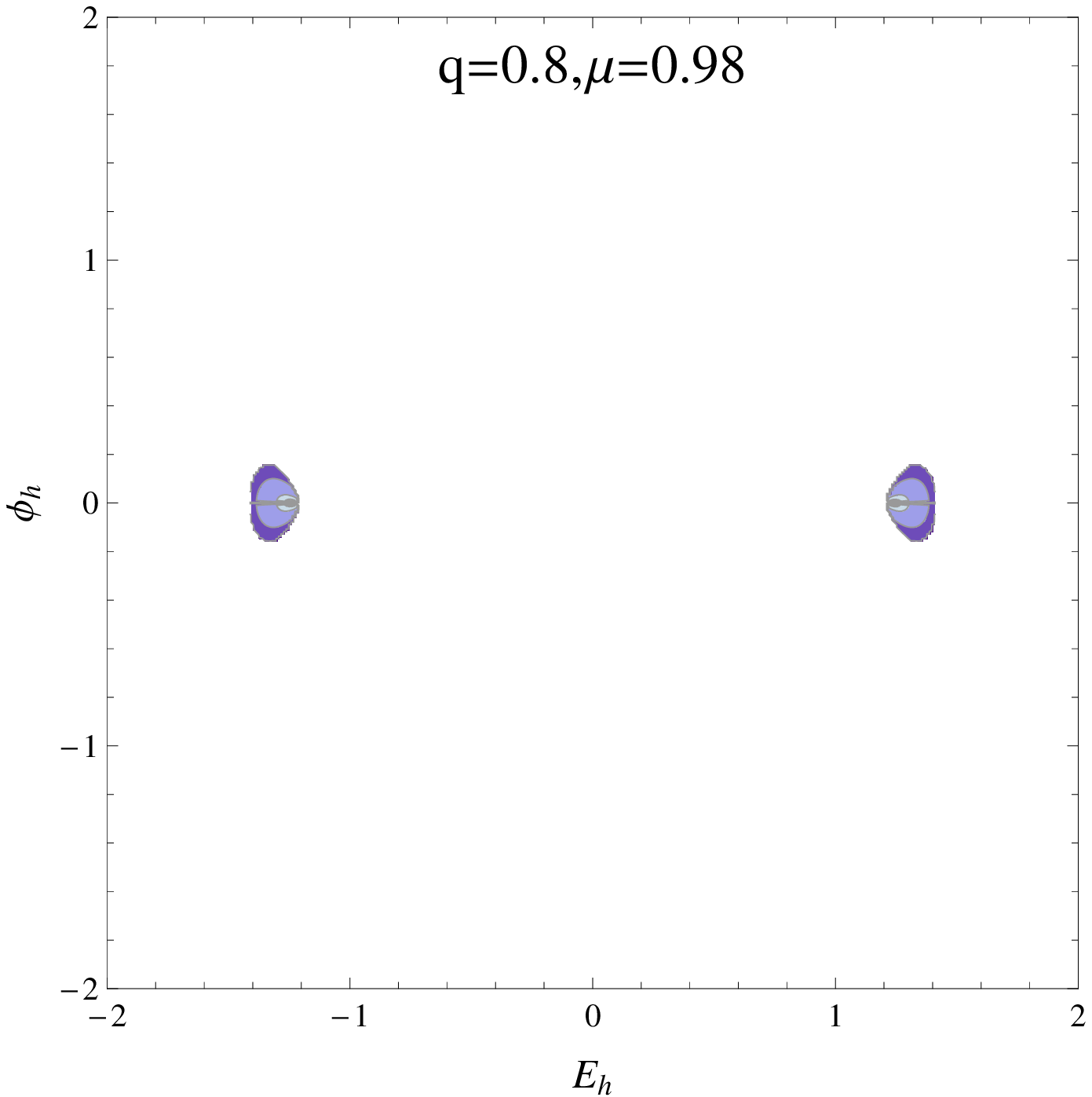}
\end{tabular}
\end{center}
\caption{Phase spaces of black hole solutions with event horizon radius $r_{h}=1$ and various values of the scalar field charge $q$ and mass $\mu $.
Shaded regions indicate where solutions exist.  The curves are contours at constant mirror radius $r_{m}=20$, $40$, $60$, $80$, $100$ and $300$, except in the last two plots ($q=0.8$, $\mu = 0.5$, $0.98$) where the outermost contour is $r_{m}=5$.
The darkest regions have $r_{m}<20$; for the lightest regions, the mirror radius $r_{m}>300$.
As the scalar field charge $q$ increases, the region containing black holes with small $r_{m}$ (the darkest blue region) increases in size.}
\label{fig:BHphase}
\end{figure*}

\subsection{Black holes}
\label{sec:BH}

We consider black holes with event horizon radius $r_{h}$, which can be set equal to unity using a length rescaling \cite{Ponglertsakul:2016wae}.
In a neighbourhood of the event horizon, the field variables have the expansions
\begin{align}
m &= \frac{r_h}{2} + m^{\prime}_h(r-r_h) + O(r-r_h)^2, \nonumber \\
h &= 1 + h^{\prime}_{h}(r-r_h) + O(r-r_h)^2, \nonumber \\
A_0 &= E_h(r-r_h) + \frac{A^{\prime\prime}_h}{2}(r-r_h)^2 + O(r-r_h)^3, \nonumber \\
\phi &= \phi_h + \phi_h' (r-r_h)   +\frac{\phi^{\prime\prime}_h}{2}(r-r_h)^2 + O(r-r_h)^3,
\label{eq:horizon}
\end{align}
where
\begin{align}
m^{\prime}_h &= \frac{r_h^2}{4}\left(\mu^2 \phi_h^2 + E^2_h\right), \qquad
h^{\prime}_h = \frac{4r_h^3\phi_h^2 \left(\mu^4 + q^2 E^2_h\right)}{\left[2-r_h^2\left(\mu^2\phi_h^2 + E^2_h\right)\right]^2}, \nonumber \\
\phi^{\prime}_h &= \frac{2r_h\mu^2\phi_h}{2 - r_h^2\left(\mu^2\phi_h^2 + E^2_h  \right)},
\label{eq:horizon1}
\end{align}
and $A_{h}''$ and $\phi _{h}''$ are given in terms of $q$, $\mu $, $r_{h}$, $\phi _{h}$ and $A_{h}'=E_{h}$.
For fixed $\mu $ and $q$, with $r_{h}=1$, black hole solutions are parameterized by $\phi _{h}$ and $E_{h}$.
In order for the event horizon to be nonextremal, we find that $E_{h}^{2}+\mu ^{2}\phi _{h}^{2}<2$ when $r_{h}=1$, which restricts the black hole phase space.

Some typical scalar field profiles for black hole solutions are shown in figure \ref{fig:BHex}.
When the scalar field is massless, $\phi _{h}'=0$ and $\phi _{h}''$ has the opposite sign to $\phi _{h}$ \cite{Dolan:2015dha}.
Therefore, for $\mu =0$ and $\phi _{h}>0$, the scalar field is decreasing close to the horizon.
For a massive scalar field,  from (\ref{eq:horizon1}) we see that $\phi _{h}'$ has the same sign as $\phi _{h}$.  Therefore, when $\phi _{h}>0$, the scalar field is increasing close to the event horizon and has a maximum between the event horizon and mirror.
This behaviour can be seen in the scalar field profiles shown in figure \ref{fig:BHex}, and in the final scalar field configurations resulting from the time-evolution of the charged black hole bomb instability \cite{Sanchis-Gual:2015lje, Sanchis-Gual:2016tcm}.

The phase spaces of black hole solutions for various values of the scalar field charge $q$ and mass $\mu $ are shown in figure \ref{fig:BHphase}.
When $\mu >0$, we find that there is a minimum value of $\left| E_{h} \right| $  for which there are nontrivial black holes. This minimum is very small when $q$ is large and $\mu $ is small, when the gap in the phase space for small $\left| E_{h} \right| $ is not visible in figure \ref{fig:BHphase}. Below this minimum, the scalar field does not have a zero before either $f(r)$ has a second zero or the solution becomes singular.

\begin{figure}
\begin{center}
\includegraphics[width=0.9\columnwidth]{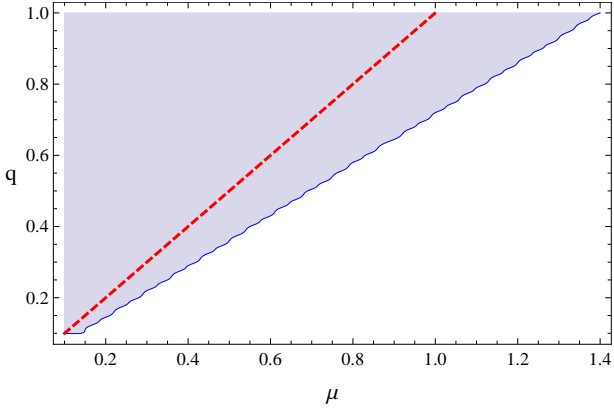}
\end{center}
\caption{Phase space of black hole solutions with event horizon radius $r_{h}=1$. The shaded region denotes those values of scalar field mass $\mu $ and charge $q$ for which we find hairy black holes.  The red dashed line is $q=\mu $.  It is clear that we find solutions for which $\mu > q$.}
\label{fig:BHqmu}
\end{figure}

For each value of the scalar field charge $q$, we find a maximum value of the scalar field mass $\mu $ for which there are hairy black hole solutions.
In figure \ref{fig:BHqmu} we plot the region of the $(q,\mu )$-plane (with event horizon radius $r_{h}=1$ and $0<q<1$) for which there are black hole solutions.
It is clear that, for each value of the scalar field charge $q$,
the maximum scalar field mass is always larger than $q$, in other words we find nontrivial black holes with $\mu >q$.

\section{Stability analysis}
\label{sec:stab}

We now examine the stability of the soliton and black hole solutions under linear, spherically symmetric, perturbations of the metric, electromagnetic field and scalar field. The method is largely unchanged from that employed in \cite{Dolan:2015dha, Ponglertsakul:2016wae} in the massless case. We therefore simply state the perturbation equations and briefly discuss the numerical results, referring the reader to \cite{Dolan:2015dha, Ponglertsakul:2016wae} for details of the derivation and numerical method used.

\subsection{Perturbation equations}
\label{sec:pert}

We begin by introducing two new field variables:
\begin{equation}
\gamma = fh^{1/2}, \qquad \psi = r\phi ,
\end{equation}
where now $\gamma $, $f$, $h$, $A_{0}$, $\phi $ and $\psi $ depend on the radial coordinate $r$ and time $t$.
We write the field variables as, for example, $f(t,r)={\bar {f}}(r)+\delta f(t,r)$ where barred variables are static equilibrium quantities and $\delta f$ (with similar notation for the other variables) are  time-dependent  perturbations.
All perturbations are real, apart from the scalar field perturbation $\delta \psi $, which we write in terms of its real and imaginary parts as \cite{Dolan:2015dha}:
\begin{equation}
\delta \psi (t,r) = \delta u(t,r) + i\delta\dot{w}(t,r),
\label{eq:deltapsi}
\end{equation}
where $\delta u$ and $\delta w$ are real.
The derivation of the linearized perturbation equations is essentially the same as in the massless case \cite{Dolan:2015dha, Ponglertsakul:2016wae}. The metric perturbations can be eliminated to give three perturbation equations for $\delta u$, $\delta w$ and $\delta A_{0}$. The final perturbation equations are slightly modified by the inclusion of the scalar field mass $\mu $, and take the form
\begin{subequations}
\label{eq:perteqns}
\begin{align}
0 = & \delta\ddot{u} - \bg^2\delta u'' -\bg\bg'\delta u' + \left[3q^2\bA^2+\frac{\bg\bg'}{r}-\barf\bh \left(\frac{\bar{\psi}}{r}\right)'^2
\right. \nonumber \\ & \left.
+\frac{\barf\bA'^2}{2}\left(\left(\frac{\bar{\psi}}{r}\right)^2+\bpsi'^2\right) -\frac{\barf\bpsi\bpsi'\bA'^2}{r}
\right. \nonumber \\ & \left.
 + \mu^2 \bar{f}\bar{h}\left\{ 1 + \bar{\psi}\left(\frac{\bar{\psi}}{r}      \right)'\left(2 + \frac{\bar{\psi}}{2} \left(\frac{\bar{\psi}}{r}      \right)' \right) \right\} \right]\delta u + 2q\bA\bg^2\delta w''
 \nonumber \\ &
 + q\barf\bA\left[2\sqrt{\bh}\bg'
 +\left(-\frac{\bA'}{\bA}\mathcal{A} +\frac{\bh}{r} + \frac{r\bA'^2}{2}\right)\left(\frac{\bar{\psi}}{r}\right)'\bpsi
 \right. \nonumber \\ & \left.
 - \frac{\mu^2\bar{h}\bar{\psi}^2}{r}\left(1 + \frac{\bar{\psi}}{2}\left(\frac{\bar{\psi}}{r}\right)'\right)\right]\delta w'
+ q\bA\left[2q^2\bA^2 -\frac{2\bg\bg'}{r}
\right. \nonumber \\ & \left.
+ \bg\bpsi'\left(\frac{\bar{\psi}}{r}\right)'\left(\frac{\bg\bA'}{\bA}-\bg'-\frac{\bg}{r}\right) + \frac{\mu^2\bar{f}\bar{h}\left(-2r + \bar{\psi}\bar{\psi}'\right)}{r}\right]\delta w ,
\end{align}
\begin{align}
0 = & \delta\ddot{w} - \bg^2\delta w'' + \left[-\bg\bg' + \frac{q^2\bA\bpsi^2}{r^2\bA'}\mathcal{A}\right]\delta w' + \left[-q^2\bA^2
\right. \nonumber \\ & \left.
- \frac{q^2\bA\bpsi\bpsi'}{r^2\bA'}\mathcal{A}
+\frac{\bg\bg'}{r} + \mu^2\bar{f}\bar{h} \right]\delta w
-q\bA\left[2  + \bpsi\left(\frac{\bar{\psi}}{r}\right)'\right]\delta u
\nonumber \\ &
+ \frac{q\bA\bpsi}{\bA'}\delta A'_0 - q\bpsi\delta A_0,
\end{align}
\begin{align}
0 = & \frac{q\bpsi}{\bA' r^2}\mathcal{A}\delta w'' + \frac{q\bpsi\bA}{r^2}\left[\frac{\bg'}{\bA\bA'\bg}\mathcal{A} - \frac{q^2\bpsi^2\bh}{r^2\bA'^2}\right]\delta w'
\nonumber \\ &
 + \frac{q\bpsi\bA}{r^2}\left[\frac{\mathcal{A}}{r\bA\bA'\bg}\left(-\bg'+\frac{rq^2\bA^2}{\bg} - \mu^2 r \sqrt{\bar{h}}\right) + \frac{q^2\bh\bpsi\bpsi'}{r^2\bA'^2} \right]\delta w
 \nonumber \\ &
  -\left(\frac{\bar{\psi}}{r}\right)'\delta u' -\left[\left(\frac{\bar{\psi}}{r}\right)''+\left(\frac{1}{r} + \frac{\bg'}{\bg}\right)\left(\frac{\bar{\psi}}{r}\right)' - \frac{\mu^2\bpsi}{r \barf}\right]\delta u + \left[\frac{\delta A'_0}{\bA'}\right]' ,
\end{align}
\end{subequations}
where we have defined
\begin{equation}
{\mathcal {A}} = \barf\bh + r\bA\bA' .
\end{equation}
At the mirror $r=r_{m}$, the scalar field perturbations $\delta u$ and $\delta w$ must vanish; there is no restriction on the value of $\delta A_{0}$ there.
The other boundary conditions depend on whether we are considering equilibrium solitons or black holes.

\subsection{Solitons}
\label{sec:solstab}

For soliton solutions,  we consider time-periodic perturbations of the form \cite{Ponglertsakul:2016wae}
\begin{align}
\delta u(t,r) & = \text{Re}\left[ e^{-i\sigma t}\tilde{u}(r) \right],
\qquad
\delta w(t,r)  = \text{Re}\left[ e^{-i\sigma t}\tilde{w}(r) \right],
\nonumber \\
\delta A_0(t,r) &  = \text{Re} \left[ e^{-i\sigma t}{\tilde{A}}_{0}(r) \right] ,
\label{eq:timepsol}
\end{align}
where ${\tilde {u}}$, ${\tilde {w}}$, ${\tilde {A}}_{0}$ have the following expansions near the origin
\begin{equation}
\tilde{u} = r \sum^{\infty}_{j=0} u_j r^j,
\qquad
\tilde{w} = r \sum^{\infty}_{j=0} w_j r^j,
\qquad
{\tilde{A}}_0 = \sum^{\infty}_{j=0} \alpha_j r^j .
\label{eq:originperts}
\end{equation}
As in \cite{Ponglertsakul:2016wae}, we can use the residual gauge and diffeomorphism freedom to set $w_{0}=0=\alpha _{0}$ and fix $u_{0}$ since the perturbation equations (\ref{eq:perteqns}) are linear.
This leaves $\sigma ^{2}$ and $w_{2}$ as free parameters.
We find that $u_{1}$, $w_{1}$, $\alpha _{1}$ all vanish and subsequent terms in the expansions (\ref{eq:originperts}) are determined by $\sigma ^{2}$, $w_{2}$ and $u_{0}$.

\begin{figure}
\begin{center}
\includegraphics[width=0.9\columnwidth]{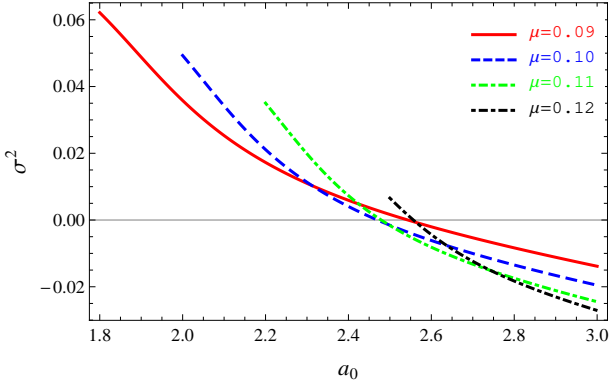}
\end{center}
\caption{Smallest eigenvalue $\sigma ^{2}$ for solitons with scalar field charge $q=0.1$ and four values of the scalar field mass $\mu $. We have fixed $\phi _{0}=1.4$.}
\label{fig:solstab}
\end{figure}

In figure \ref{fig:solstab} we plot the smallest eigenvalue $\sigma ^{2}$ (which we find to be real) for some typical soliton solutions.  The results are very similar to those found in \cite{Ponglertsakul:2016wae} when the scalar field mass $\mu =0$.
Although including a scalar field mass $\mu $ does change the numerical values of the eigenvalues $\sigma ^{2}$, the qualitative results from \cite{Ponglertsakul:2016wae} are unchanged.
In particular, for larger values of the mirror radius, all soliton solutions we investigated have $\sigma ^{2}>0$, so that the perturbation frequency $\sigma $ is real and the solutions are stable.
However, if the mirror radius is sufficiently small, then we find that some solitons have eigenvalues $\sigma ^{2}<0$, giving a purely imaginary perturbation frequency.  In this case there are perturbations which grow exponentially with time and hence the solitons are unstable.
When $\mu >q$, we still find both stable and unstable solitons.

\subsection{Black holes}
\label{sec:BHstab}

\begin{figure}
\begin{center}
\includegraphics[width=0.9\columnwidth]{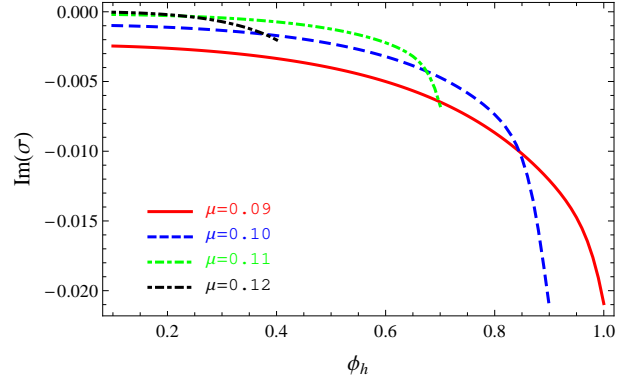}
\end{center}
\caption{Imaginary part of the perturbation frequency $\sigma $ for black hole solutions with scalar field charge $q=0.1$ and four values of the scalar field mass.  We have fixed $E_{h}=1.2$ and the event horizon radius $r_{h}=1$.}
\label{fig:BHstab}
\end{figure}

Perturbations of black hole solutions have ingoing boundary conditions at the event horizon, so we consider \cite{Dolan:2015dha}:
\begin{align}
\delta u(t,r) &= \text{Re}\left[ e^{-i\sigma (t+r_{*})}\tilde{u}(r) \right] ,
\quad
\delta w(t,r) = \text{Re} \left[ e^{-i\sigma (t+r_{*})}\tilde{w}(r) \right] ,
\nonumber \\
\delta A_0(t,r) &= \text{Re} \left[ e^{-i\sigma (t+r_{*})}{\tilde{A}}_{0}(r) \right] ,
\label{eq:BHperts}
\end{align}
where the usual tortoise coordinate $r_{*}$ is defined by
\begin{equation}
\frac {dr_{*}}{dr} = \frac{1}{\bg} .
\end{equation}
The quantities ${\tilde {u}}$, ${\tilde {w}}$ and ${\tilde {A}}_{0}$ have the following expansions near the horizon:
\begin{align}
\tilde{u} &= {\tilde {u}}_{0} + {\tilde {u}}_{1}(r-r_h) + O(r-r_h)^2, \nonumber \\
\tilde{w} &= {\tilde {w}}_{0} + {\tilde {w}}_{1}(r-r_h) + O(r-r_h)^2, \nonumber \\
{\tilde{A}}_0 &= {\tilde{A}}_{1}(r-r_h) + {\tilde{A}}_{2}(r-r_h)^2 + O(r-r_h)^3.
\label{eq:BHperts1}
\end{align}
Since the perturbation equations are linear, we can fix ${\tilde {u}}_{0}$ without loss of generality, and then ${\tilde {u}}_{1}$, ${\tilde {w}}_{1}$, ${\tilde {A}}_{1}$ and subsequent terms in the expansions (\ref{eq:BHperts1}) are determined by ${\tilde {w}}_{0}$ and the eigenvalue $\sigma $.

In contrast to the soliton case, for equilibrium black hole solutions the eigenvalue $\sigma $ is, in general, complex. In figure \ref{fig:BHstab} we show the imaginary part of $\sigma $ for some typical black hole solutions.
Again our results are qualitatively similar to those obtained in \cite{Dolan:2015dha} when $\mu =0$, although the numerical values of $\sigma $ depend on the scalar field mass.
In particular, for all the black holes we investigated (including those with $\mu >q$), we find that the imaginary part of $\sigma $ is negative, so the perturbations (\ref{eq:BHperts}) are exponentially decaying with time and the black holes are stable.

\section{Conclusions}
\label{sec:conc}

We have studied the effect of introducing a scalar field mass $\mu $ on static, spherically symmetric, charged scalar solitons and black holes in a cavity, studied for $\mu =0$ in \cite{Dolan:2015dha, Ponglertsakul:2016wae}.
For black hole solutions, we find that the scalar field must have a maximum outside the event horizon if it is positive on the horizon.
For solitons, if the scalar field is positive at the origin, it may have a maximum either at the origin, or between the origin and the reflecting mirror at $r=r_{m}$.

The phase spaces of soliton and black hole solutions have a number of interesting new features when $\mu $ is nonzero.
For fixed scalar field charge $q$, for both solitons and black holes the phase space shrinks as $\mu $ increases, with a nonzero lower bound on the magnitude of either the electrostatic potential at the origin (for solitons) or the derivative of the electrostatic potential at the horizon (for black holes).
For black hole solutions, for fixed $q$ there is a maximum value of the scalar field mass $\mu $ for which we find solutions.

We have also studied the dynamical stability of our solutions under linear, spherically symmetric perturbations of the metric, scalar field and electromagnetic field.  Recently, the thermodynamic stability of solitons and hairy black holes with a massless charged scalar field in a cavity has been studied \cite{Basu:2016srp}. A complex thermodynamic phase space emerges, in some regions of which the solitons or the hairy black holes are the thermodynamically stable configuration.  It would be interesting to investigate the effect of a scalar field mass $\mu $ on the thermodynamic phase space.

Our work was motivated by the question of the end-point of the charged black hole bomb instability, which occurs in the test-field limit if the scalar field mass $\mu $ and charge $q$ satisfy the inequality $q>\mu $ \cite{Herdeiro:2013pia, Hod:2016nsr}.  The hairy black holes we find with $q>\mu >0$ are possible end-points of this instability. When the mirror is located at the first zero of the scalar field, the hairy black holes appear to be linearly stable. Furthermore, the static equilibrium solutions we find here are identical (after a gauge transformation) to the final black hole configurations found in \cite{Sanchis-Gual:2015lje, Sanchis-Gual:2016tcm} from a time-evolution of a Reissner-Nordstr\"om black hole in a cavity with a charged scalar field perturbation.
The fact that we have a lower bound on $\left| E_{h} \right|$ for fixed $\mu $ and $q$ for hairy black hole solutions sets a limit on the amount of charge that the scalar field can extract from the black hole during the evolution of the charged black hole bomb (see \cite{Sanchis-Gual:2015lje, Sanchis-Gual:2016tcm} for detailed studies of the extraction of charge and energy from the black hole as the charged black hole bomb evolves).

In this context our solutions with $\mu >q$ are particularly interesting.
When $\mu >q$, a linearized probe charged scalar field on a Reissner-Nordstr\"om black hole background does not exhibit a charged black hole bomb instability \cite{Herdeiro:2013pia, Hod:2016nsr}.
Since we find both soliton and black hole solutions with $\mu >q$, we can nonetheless interpret the hairy black holes as bound states of the solitons and a bald Reissner-Nordstr\"om black hole.  We conjecture that the black holes in this case could form from the gravitational collapse of an unstable soliton with $\mu >q$.  To test this conjecture, a full nonlinear time-evolution of the Einstein-Maxwell-Klein-Gordon equations would be required, which we leave for future work.

\par
{\it {Note added:}}
Very recently, the evolution of unstable solitons when the charged scalar field mass $\mu =0$ has been studied
\cite{Sanchis-Gual:2016ros}.  It is found that a black hole forms, which is either a bald Reissner-Nordstr\"om black hole or can have nontrivial charged scalar field hair.
It would be interesting to extend the investigation of \cite{Sanchis-Gual:2016ros} to include a nonzero scalar field mass $\mu $.

\section*{Acknowledgments}
We thank Carlos Herdeiro for insightful discussions and Shahar Hod for helpful comments. The work of EW is supported by the Lancaster-Manchester-Sheffield Consortium for
Fundamental Physics under STFC grant ST/L000520/1.


\end{document}